\documentclass{article}

\PassOptionsToPackage{numbers, compress}{natbib}
\usepackage[preprint]{neurips_2024}

\usepackage[utf8]{inputenc} 
\usepackage[T1]{fontenc}    
\usepackage[colorlinks,
            linkcolor=red,
            anchorcolor=blue,
            citecolor=green
            ]{hyperref}
\usepackage{url}            
\usepackage{booktabs}       
\usepackage{fontawesome}   
\usepackage{amsfonts}       
\usepackage{nicefrac}       
\usepackage{microtype}      
\usepackage{xcolor}         
\usepackage{caption}
\usepackage{subcaption}
\usepackage{natbib}
\usepackage{graphicx}
\usepackage{enumitem}
\usepackage{makecell}
\usepackage{amsmath}
\usepackage{amsthm}
\usepackage{thmtools}
\usepackage{thm-restate}
\usepackage{algorithm}
\usepackage{algorithmic}
\usepackage{bbm}
\usepackage{tabularx}
\usepackage{listings}
\usepackage{xcolor}
\usepackage{multirow}
\usepackage{multicol}
\usepackage{tcolorbox}
\usepackage{tikz}
\usepackage{footnote}
\usepackage{wrapfig}

\usepackage{multirow}
\usepackage{float}      

\definecolor{tblue}{RGB}{31,119,180}
\definecolor{torange}{RGB}{255,127,14}
\definecolor{tgreen}{RGB}{44,160,44}
\definecolor{tred}{RGB}{214,39,40}
\definecolor{tpurple}{RGB}{148,103,189}
\definecolor{lightblue}{RGB}{173, 216, 230}
\definecolor{lightpink}{RGB}{255, 182, 193}
\definecolor{lightgreen}{RGB}{144, 238, 144}

\usepackage{colortbl}
\usepackage{xcolor}
\usepackage{array}

\newcommand{\hide}[1]{} 

\newcommand{\eg}{\textit{e}.\textit{g}.}

\def\model{RecGPT}

\title{RecGPT: A Foundation Model for Sequential Recommendation}

\author{
  Yangqin Jiang\textsuperscript{1} ~~~
  Xubin Ren\textsuperscript{1} ~~~
  Lianghao Xia\textsuperscript{1} ~~~
  Da Luo\textsuperscript{2} ~~~
  Kangyi Lin\textsuperscript{2} ~~~
  Chao Huang\textsuperscript{1}\thanks{Chao Huang is the Corresponding Author.} \\
  \textsuperscript{1}The University of Hong Kong ~~~
  \textsuperscript{2}Tencent. \\
  \texttt{\{mrjiangyq99, xubinrencs, chaohuang75\}@gmail.com, aka\_xia@foxmail.com} \\
  \faGithub~\textbf{Source Code:} \textcolor{blue}{\url{https://github.com/HKUDS/\model}
}}

\begin{document}

\maketitle

\begin{abstract}
This work addresses a fundamental barrier in recommender systems: the inability to generalize across domains without extensive retraining. Traditional ID-based approaches fail entirely in cold-start and cross-domain scenarios where new users or items lack sufficient interaction history. Inspired by foundation models' cross-domain success, we develop a foundation model for sequential recommendation that achieves genuine zero-shot generalization capabilities. Our approach fundamentally departs from existing ID-based methods by deriving item representations exclusively from textual features. This enables immediate embedding of any new item without model retraining. We introduce unified item tokenization with Finite Scalar Quantization that transforms heterogeneous textual descriptions into standardized discrete tokens. This eliminates domain barriers that plague existing systems. Additionally, the framework features hybrid bidirectional-causal attention that captures both intra-item token coherence and inter-item sequential dependencies. An efficient catalog-aware beam search decoder enables real-time token-to-item mapping. Unlike conventional approaches confined to their training domains, \model\ naturally bridges diverse recommendation contexts through its domain-invariant tokenization mechanism. Comprehensive evaluations across six datasets and industrial scenarios demonstrate consistent performance advantages.
\end{abstract}


\section{Introduction}\label{sec:intro}

Recommender systems (RSs) have emerged as indispensable tools for navigating the overwhelming sea of digital content, offering personalized guidance across diverse platforms including e-commerce marketplaces~\cite{wang2020time}, multimedia streaming services~\cite{jiang2024diffmm}, and social networking sites~\cite{jamali2010matrix}. By intelligently filtering vast information landscapes, these systems not only enhance user satisfaction but also drive engagement and retention for platform operators. Within this ecosystem, sequential recommendation approaches have gained particular prominence for their ability to capture temporal dynamics and evolving preferences, enabling more accurate predictions of users' future interactions by modeling the intricate patterns within their historical behavior sequences~\cite{fang2020deep}.

In the sequential recommendation, existing frameworks hit fundamental barriers when confronted with new contexts, data-sparse environments, or cold-start scenarios~\cite{zhao2023cross,zhang2025cold}, invariably demanding resource-intensive retraining cycles. This critical limitation creates an urgent need for a transformative recommendation paradigm with robust \textbf{Zero-Shot Generalization} capabilities that can adapt seamlessly across diverse scenarios without prior exposure to specific users or items, effectively making recommendations in previously unseen contexts while maintaining recommendation quality.

\begin{wrapfigure}{r}{0.55\textwidth}
    \centering
    \vspace{-0.2in}
    \includegraphics[width=0.52\textwidth]{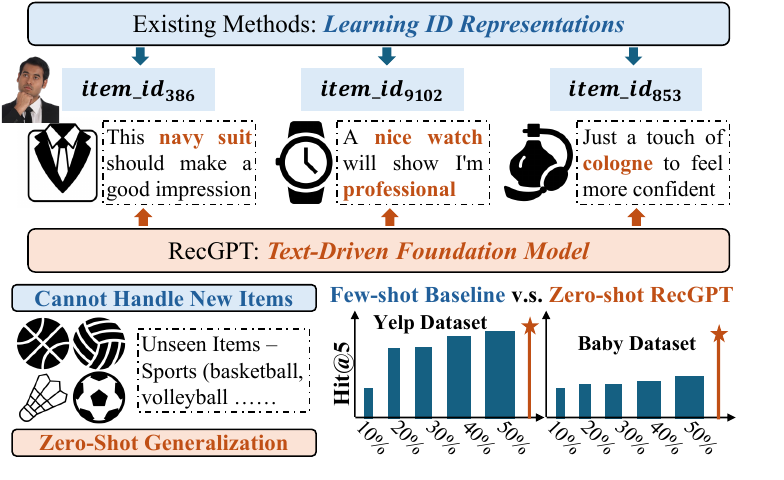}
    \caption{Our \model\ model demonstrates strong cross-domain zero-shot generalization capabilities, consistently outperforming existing recommender systems in few-shot scenarios (even when those systems incorporate 10\%-50\% of downstream unseen data) without requiring any domain-specific training data.}
    \label{fig:intro}
    \vspace{-0.15in}
\end{wrapfigure}

Inspired by the remarkable success of foundation models in visual and language domains~\cite{liu2023llava,deepseekv3}, which achieve exceptional cross-domain generalization through large-scale pre-training, a compelling question emerges: \textbf{Can we develop foundation models for sequential recommenders with effective pre-training paradigms?} Such models would enable efficient knowledge transfer across diverse recommendation scenarios without the computational burden of extensive retraining. This capability would be particularly valuable in challenging sparse data conditions, cold-start situations, and zero-shot recommendation contexts that current recommender systems struggle to address effectively. However, realizing this vision requires addressing several fundamental challenges:

\textbf{The Semantic Heterogeneity of Item Representations}. Inspired by the success of autoregressive LLMs in achieving cross-domain generalization through next-token prediction, we recognize a natural fit for sequential recommendation, which inherently models temporal patterns in user interactions. However, directly applying this paradigm faces a critical challenge: the semantic heterogeneity of item representations across domains. Unlike language tokens that share universal linguistic structures, items from different platforms (\eg, e-commerce products, videos, news) exhibit vastly different descriptive formats and attribute spaces. This heterogeneity creates insurmountable barriers for traditional ID-based systems, which fail to transfer knowledge across domains. To address this limitation, recommendation foundation models must develop a domain-invariant tokenization mechanism that unifies diverse textual descriptions into a standardized discrete token space--preserving semantic richness while enabling the powerful generalization capabilities of autoregressive modeling.

\textbf{Hierarchical Dependencies in Item Tokenization}. Language tokenization enables straightforward autoregressive modeling—once words are mapped to tokens, the model simply predicts the next token without explicit word-level considerations. Item tokenization, however, presents a fundamentally different challenge: when an item is represented by multiple tokens, the model must simultaneously capture two levels of dependencies. At the macro level, it must model sequential relationships between items to understand user preference evolution. At the micro level, it must maintain coherence among tokens belonging to the same item to preserve its semantic integrity. This dual-dependency structure demands a more sophisticated attention mechanism that can distinguish between intra-item and inter-item relationships while maintaining the autoregressive framework necessary for generalization.

\textbf{The Decoding Efficiency Challenge}. The computational complexity of converting token predictions to item recommendations creates a critical bottleneck for real-world deployment. The vast space of possible token combinations far exceeds actual catalog sizes, making naive search approaches computationally prohibitive—a vocabulary of 50,000 tokens with 4-token items yields $6.25\times 10^18$ theoretical combinations versus millions of actual items. Therefore, efficient decoding mechanisms must exploit semantic structure and approximate search techniques to map token predictions to valid items at inference speeds compatible with real-time recommendation systems.

To address these challenges, we present \model, a text-driven foundation model that reimagines sequential recommendation through the lens of autoregressive generation. Our framework introduces three architectural designs that collectively enable zero-shot generalization capabilities. First, we develop a Unified Item Tokenization mechanism that employs Finite Scalar Quantization (FSQ) to transform heterogeneous textual descriptions into a standardized discrete token space. This domain-invariant representation solves the semantic heterogeneity problem by creating a universal vocabulary that bridges diverse recommendation contexts, without requiring platform-specific adaptations. Unlike traditional ID-based systems confined to their training domains, our tokenization preserves rich semantic information while enabling cross-domain knowledge transfer.

Second, we propose a Universal Recommendation Modeling architecture featuring hybrid bidirectional-causal attention that elegantly resolves the hierarchical dependency challenge inherent in multi-token item representations. This innovative attention mechanism allows comprehensive information exchange among tokens representing the same item through bidirectional processing, while maintaining strict causal relationships between sequential items for accurate temporal modeling. We further introduce auxiliary semantic pathways that complement discrete tokens with continuous embeddings, effectively counteracting information loss during quantization. Third, our Efficient Item Token Decoder employs catalog-aware beam search with Trie-based prefix constraints, transforming the computationally intractable token-to-item mapping into a practical real-time operation. By exploiting the sparse nature of valid item combinations within the vast token space, our decoder achieves optimal complexity while maintaining superior recommendation quality.

Through experiments on several datasets and industrial deployment serving millions of users, we show that \model\ achieves superior zero-shot performance--consistently outperforming traditional recommenders. Our comprehensive analysis reveals exceptional cold-start capabilities, robust power-law scaling properties, and architectural advantages confirmed through systematic ablations.

\section{The \model\ Framework}

We propose a \textbf{Text-Driven Foundation Model} for sequential recommendation that fundamentally departs from traditional ID-based approaches. While conventional recommenders learn separate embeddings for each unique item ID--creating inherent limitations in generalization capabilities and domain adaptability~\cite{jiang2024reclm}--our approach derives item representations exclusively from textual features (\textit{e.g.}, titles, descriptions, and categories) through a specialized text encoder. This paradigm shift offers three substantial advantages: (1) \textbf{Zero-shot Transferability}, as any new item can be immediately embedded via its textual description without model retraining; (2) \textbf{Cross-domain Compatibility}, since textual semantics naturally bridge diverse recommendation contexts ranging from e-commerce products to video content and news articles; and (3) \textbf{Enhanced Robustness} in sparse-data and cold-start scenarios, where ID-based methods typically fail due to insufficient interaction histories but text-based representations remain informative through their rich semantic content.

\subsection{Unified Item Tokenization}
Our unified item tokenization framework addresses a critical challenge in cross-domain recommendation: semantic heterogeneity of item descriptions. By transforming diverse textual representations into standardized tokens, we eliminate domain barriers that hinder traditional recommenders. The sequence encoder $\Phi(\mathcal{S}) = g_\phi([f_\theta(\mathbf{x}1), ..., f_\theta(\mathbf{x}_n)])$ leverages large-scale data across multiple domains, capturing universal interaction patterns that transcend specific platforms or categories. This multi-domain pre-training develops genuine zero-shot capabilities absent in conventional systems.

The tokenization process first embeds items into a consistent representation space, then employs our novel quantization mechanism with a \textbf{Unified Codebook} to convert these continuous embeddings into discrete tokens. Unlike previous approaches that suffer from information loss during quantization, our strategy preserves essential semantic relationships while enabling efficient autoregressive modeling. This balanced approach maintains representational fidelity while supporting scalable pre-training, addressing the fundamental trade-off between computational efficiency and recommendation quality. Consequently, our model can generate highly personalized recommendations for completely unseen items or domains without requiring domain-specific fine-tuning or retraining—a capability unattainable with traditional ID-based recommenders.

\subsubsection{Text-based Item Representation}
We leverage MPNet~\cite{song2020mpnet} as our foundational text encoder to derive item representations from textual descriptions. This strategic choice capitalizes on MPNet's innovative architecture that seamlessly integrates Masked Language Modeling (MLM) and Permuted Language Modeling (PLM), addressing fundamental limitations in contextual understanding that plague earlier transformer models~\cite{devlin2018bert, yang2019xlnet}. Unlike ID-based methods that fail to generalize to new items, our text-driven approach enables zero-shot recommendations and eliminates the need for extensive item-specific training data, particularly advantageous for cold-start scenarios and long-tail items.

For a text sequence \(X = (x_1, \ldots, x_n)\), our text encoder optimizes the following objective function:
\begin{equation}
    \mathbb{E}_{z \in \mathcal{Z}_n} \sum_{t=c+1}^{n} \text{log}P(x_{z_t} | x_{z_{<t}}, \Phi_{z_{>c}}; \theta),
\end{equation}
where $\mathcal{Z}_n$ represents permutations of indices $(1,\cdots,n)$, $z_t$ and $z{<t}$ denote the $t$-th index and first $t-1$ elements in permutation $z$, and $\Phi_{z_{>c}}$ indicates masks in positions $z_{>c}$. After training with this objective, \model\ generates $d_{L}$-dimensional embeddings $e_i$ for item $i$ with textual features $X_i$ as:
\begin{equation}
e_i = \text{MPNet}(X_i), e_i \in \mathbb{R}^{d_{L}}.
\end{equation}
This approach produces semantically rich, domain-agnostic representations that maintain consistent meaning across diverse recommendation contexts. By grounding recommendations in language encoding, our model naturally bridges domain gaps and supports cross-domain transfer without requiring explicit adaptation mechanisms or specialized embeddings for recommendation contexts.

\begin{figure*}[t]
    \centering
    \includegraphics[width=1 \linewidth]{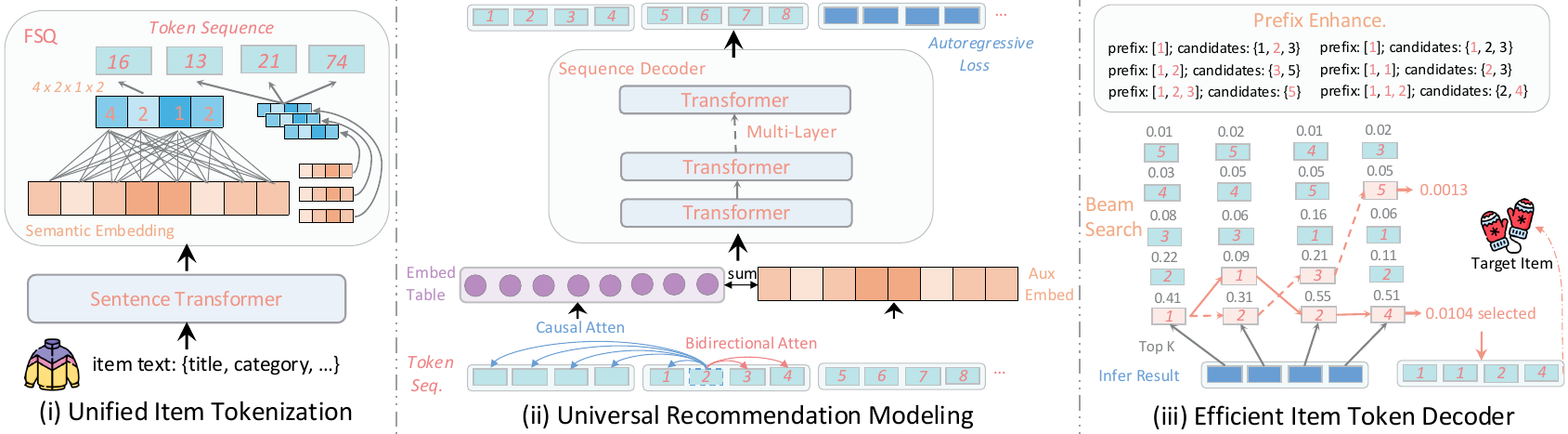}
    \vspace{-0.1in}
    \caption{Architecture of our proposed foundation model-\model\ for recommender systems, featuring three key components: (i) \textbf{Unified Item Tokenization} that converts text descriptions into discrete tokens via Sentence Transformer and FSQ quantization, (ii) \textbf{Universal Recommendation Modeling} with a transformer-based sequence encoder utilizing dual attention mechanisms, and (iii) \textbf{Efficient Token Decoder} employing enhanced beam search for cross-domain next-item prediction.}
    \label{fig:figure_overall}
    \vspace{-0.1in}
\end{figure*}

\subsubsection{Quantizing Item Embeddings}
To bridge the gap between continuous semantic spaces and discrete token-based processing, we introduce a novel embedding quantization mechanism inspired by advances in image processing~\cite{esser2021taming}. This critical component transforms our continuous item representations into discrete tokens—a representation format ideally suited for transformer architectures while maintaining semantic fidelity. We adopt Finite Scalar Quantization~\cite{mentzer2023finite} (FSQ), which elegantly resolves the codebook collapse challenges that have limited previous quantization approaches.

For each item $i$, our quantization strategy maps its embedding to a fixed-length token sequence $s_i = {s_i^{0}, s_i^{1}, \cdots, s_i^{K-1}}$, where each token $s_i^{k} \in \mathcal{C}$ belongs to a finite codebook. This transformation begins by partitioning the $d_L$-dimensional semantic embedding $e_i$ into $K$ sub-vectors $e_i^k \in \mathbb{R}^{d_L/K}$. This partitioning preserves local semantic structures while enabling more granular quantization.

The quantization process employs a carefully designed pipeline: first applying a sigmoid function $\sigma(\cdot)$ to normalize values to $(0,1)$, then using a rounding function $R[\cdot]$ with hyperparameter $L$ to map each component to one of $L$ distinct integers. To achieve sufficient representational capacity, we divide each sub-vector $e_i^k$ into $d_{fsq}$ segments, creating a vast discrete space of $L^{d_{fsq}}$ possible representations. This exponential expansion of the codebook size ($|\mathcal{C}| = L^{d_{fsq}}$) enables near-lossless mapping of semantic information while maintaining discrete token structure. Formally:
\begin{align}
\text{FSQ}(e_i^{k}) = R[(L - 1)\sigma (\mathcal{T}_{in}(e_i^{k}))],
\end{align}
where the output $\text{FSQ}(e_i^{k}) \in {0, \cdots, L-1}^{d_{fsq}}$ constrains each dimension to integers between $0$ and $L-1$. The transformation $\mathcal{T}{\text{in}}(e_i^{k}) = W{\text{in}} e_i^{k} + b$ reduces dimensionality from $d_{L}/K$ to $d_{\text{fsq}}$. This mapping requires optimizing only the linear transformation parameters, ensuring computational efficiency. To address the gradient-blocking nature of the rounding function, we implement the Straight-Through Estimator (STE) technique:
\begin{align}
        &\text{FSQ}(e_i^{k}) = (L-1)\sigma (\mathcal{T}_{in}(e_i^{k})) + \\
        \text{sg}[R[(&L-1)\sigma(\mathcal{T}_{in}(e_i^{k}))]-(L-1)\sigma(\mathcal{T}_{in}(e_i^{k}))],\nonumber
\end{align}
where sg[$\cdot$] denotes the ``stop gradient'' operation that allows gradient flow through the non-differentiable rounding function. This quantization mechanism transforms user interaction histories into universal token sequences rather than dataset-specific item IDs—a crucial advantage that enables cross-domain generalization. By operating in a shared, discrete token space, our model can seamlessly transfer knowledge across recommendation domains, significantly enhancing zero-shot capabilities and cold-start performance compared to traditional ID-based approaches.

\subsubsection{Optimization for Item Quantization}
Training our quantization function requires a reconstruction mechanism capable of restoring the original semantic fidelity from discrete tokens. We implement a sophisticated multi-layer transformer decoder that leverages bidirectional attention to capture complex dependencies between quantized sub-vectors. This architecture choice is critical—unlike simpler feed-forward approaches, transformers excel at contextualizing relationships across the entire token sequence $[\hat{e}_i^{0}, \cdots, \hat{e}_i^{K-1}]$, enabling high-fidelity reconstruction of the original embedding $\hat{e}_i$.

Our optimization strategy employs $L_1$ loss, which prioritizes sparse, robust reconstructions while being less sensitive to outliers than squared-error alternatives. This enhances preservation of distinguishing semantic features during the quantization-reconstruction process. Formally, we minimize:
\begin{align}
    \mathcal{L}_{fsq} = \sum_{i=0}^{N-1}&\| e_i - \text{Decoder}(\\
    &[\mathcal{T}_{out}(\hat{e}_i^{0}), \cdots, \mathcal{T}_{out}(\hat{e}_i^{K-1})])\|_{1},\nonumber
\end{align}
where $\hat{e}_i^{k}$ represents the quantized representation derived from $e_i^{k}$ (specifically, $\hat{e}i^{k} = \mathcal{T}_{in}(e_i^{k})$), and $\mathcal{T}_{out}$ functions as a dimensionality-expansion transform that maps each compact token from the lower-dimensional space $d_{fsq}$ back to the richer representational capacity of $d_{L}/K$.

\subsection{Universal Recommendation Modeling}
Our approach harnesses the transformative capabilities of autoregressive (AR) transformer architectures—models that have revolutionized zero-shot generalization across language~\cite{brown2020language}, vision~\cite{tian2024visual}, and multimodal domains~\cite{lu2022unified}. By reformulating sequential recommendation as next-token prediction, we unlock the remarkable generalization potential of the autoregressive paradigm for cross-domain recommendation with minimal adaptation requirements while addressing two fundamental challenges:

\textbf{i) Limitations of unidirectional attention.} Since each item in our framework is represented by a token sequence, standard unidirectional attention constrains information flow between tokens belonging to the same item. This artificial barrier significantly hampers the model's ability to form coherent item representations and accurately model user preferences for next-item prediction.

\textbf{ii) The information compression inherent in quantization.} While our quantization approach creates a sufficiently expressive discrete space to uniquely represent items, the compression process inevitably sacrifices some semantic nuances from the original item descriptions. This information bottleneck, if left unaddressed, could degrade user preference modeling accuracy.

As depicted in Fig. 1 (ii), \model\ implements two innovative architectural solutions - specifically designed to maximize information retention and flow - to overcome these fundamental challenges:

\subsubsection{Bidirectional Attention for Item Tokens}
\label{subsec:bi_atten}
To address the first challenge, \model\ implements a hybrid attention mechanism combining unidirectional causal attention for modeling sequential item relationships with bidirectional attention for tokens within each item. This architectural innovation enables comprehensive information sharing among tokens representing the same item while preserving temporal causality necessary for next-item prediction by restricting cross-item attention to previously encountered items in the sequence.

\subsubsection{Auxiliary Item Semantic Features}
To counteract potential information loss from quantization, \model\ incorporates the original semantic representations alongside the learned discrete token embeddings within the transformer architecture. For each item \(i\), we utilize a feature sequence \(\{e_i^{0}, \ldots, e_i^{K-1}\}\) where \(e_i^{k} \in \mathbb{R}^{d_L/K}\). These features are projected through a linear transformation to match the autoregressive model's dimensionality \(d_{\text{ar}}\). Leveraging the one-to-one correspondence between feature and token sequences, we generate two parallel embedding streams: auxiliary embeddings \(E_{\text{aux}} \in \mathbb{R}^{T \times d_{\text{ar}}}\) that preserve continuous semantic information, and token embeddings \(E_{\text{wte}} \in \mathbb{R}^{T \times d_{\text{ar}}}\) that capture discrete categorical properties, where \(T\) represents the maximum sequence length.

To ensure numerical stability and facilitate more effective gradient flow between these complementary embedding streams, we apply Layer Normalization (\(\text{LNorm}(\cdot)\)) independently to both embeddings. These normalized representations are then combined with positional embeddings \(E_{\text{wpe}} \in \mathbb{R}^{T \times d_{\text{ar}}}\) to produce the final input representations \(X \in \mathbb{R}^{T \times d_{\text{ar}}}\) for the transformer layers:
\begin{equation}
    X = \text{LNorm}(E_{aux}) + \text{LNorm}(E_{wte}) + E_{wpe}.
\end{equation}

\subsection{Training Objective}
\label{subsec:model_training}
Our autoregressive framework is fundamentally designed to optimize sequential prediction—a perfect alignment with the core objective of sequential recommendation systems. By framing next-item prediction as next-token forecasting, we create a natural bridge between language modeling and recommendation tasks. This approach allows us to leverage the remarkable capabilities of autoregressive transformers while tailoring them specifically for user preference modeling. Formally, we minimize the negative log-likelihood loss:
\begin{equation}
    \mathcal{L}_{\text{ar}} = -\sum_{t=0}^{T-1} \log P\left(Y_t \mid X_{<\left\lfloor \frac{t}{K} \right\rfloor \times K}\right),
\end{equation}
where $Y_t$ represents the target token at position $t$, and $X_{<\left\lfloor \frac{t}{K} \right\rfloor \times K}$ encompasses all contextual information available before position $\left\lfloor \frac{t}{K} \right\rfloor \times K$. The floor function $\left\lfloor \cdot \right\rfloor$ ensures proper item boundary maintenance during prediction. Crucially, our bidirectional attention mechanism enables the model to effectively capture intra-item dependencies, allowing simultaneous prediction of the next $K$ tokens representing a complete item representation during training.

\subsection{Efficient Item Token Decoder}
Converting token-level predictions into actionable item recommendations presents a fundamental challenge in our framework - a process that must be both computationally efficient and semantically accurate to deliver real-time, high-quality recommendations. Our item token decoder (Fig. 1 (iii)) tackles this critical bottleneck through carefully designed search algorithms that intelligently navigate the vast space of possible token combinations, effectively bridging the gap between our model's internal token representation and the discrete item catalog required for end-user recommendations.

\subsubsection{Efficient Next-item Prediction}
\model\ employs beam search to identify the top-$n$ most probable token sequences, striking an optimal balance between search space coverage and computational efficiency. A key advantage of our architecture is its ability to simultaneously predict all $K$ tokens representing the next item during a single inference pass, eliminating the need for iterative token generation. This parallel prediction capability allows us to directly perform beam search on the joint probability distribution of these $K$ tokens, dramatically reducing both inference latency and memory consumption compared to traditional token-by-token generation approaches.

\subsubsection{Catalog-aware Search Optimization}
A critical insight driving our search optimization is the significant disparity between the theoretical token sequence space ($L^{d_{\text{fsq}}}$) and the actual number of valid items in recommendation catalogs. Even with millions of catalog items, this represents only a minute fraction of all possible token combinations. Capitalizing on this observation, we implement a Trie-based prefix matching system that encodes the complete token vocabulary of the item catalog.

This specialized data structure enables powerful constraint-guided exploration during beam search—as each token is generated, we traverse the Trie to identify only those branches leading to valid catalog items. This approach offers two substantial benefits: (1) it prevents wasted computation on token sequences that could never yield valid recommendations, and (2) it guarantees that our final recommendations will always correspond to actual catalog items.

\section{Evaluation}
\subsection{Experimental Setup}
\noindent \textbf{Comprehensive Multi-Domain Evaluation Framework}. To rigorously evaluate \model's effectiveness across diverse recommendation scenarios, we conduct extensive experiments on six datasets spanning multiple domains and interaction patterns. These benchmarks provide varying levels of sequence complexity, interaction sparsity, and catalog size, allowing us to comprehensively assess our model's versatility and robustness under different recommendation challenges. Our evaluation framework follows established protocols from recent literature~\cite{qiu2022contrastive}, including standardized train-validation-test splits and metrics focused on ranking quality. Full dataset statistics and preprocessing details are provided in Appendix~\ref{app:data}, with evaluation protocols elaborated in Appendix~\ref{app:evaluation_setting}.

\noindent \textbf{Comparative Baseline Selection}. For rigorous analysis, we benchmark \model\ against state-of-the-art recommender systems representing distinct methodological streams. This diverse selection ensures our evaluation captures the landscape of contemporary recommendation techniques rather than narrowly comparing against a single methodology. Comprehensive baseline descriptions and implementation details, including our architecture specifications, training procedures, and parameter settings, are documented in Appendices~\ref{app:baselines} and~\ref{app:implementation}.

\subsection{Cross-domain Zero-shot Recommendation}
\label{sec:exp_zero_shot}
\begin{table*}[h]
    \centering
    \caption{Zero-shot performance of \model (no target domain training) versus Few-shot performance of competitive baselines (with 10\% target domain data access). Results clearly demonstrate \textbf{bold} (best) and \underline{underlined} (second-best) performance metrics across evaluation criteria. Statistical significance (p < 0.05) indicated by * throughout comprehensive experimental results.}
    \vspace{-0.05in}
    \setlength{\tabcolsep}{0.15mm}
    \scriptsize
    {
    \begin{tabular}{c| l| c c| c c| c c| c c| c c| cc|cc|cc|cc|cc}
        \toprule
         \multirow{2}{*}{Data} & \multirow{2}{*}{Metric} & \multicolumn{2}{c|}{GRU4Rec} & \multicolumn{2}{c|}{GRU4RecF} & \multicolumn{2}{c|}{Caser} & \multicolumn{2}{c|}{BERT4Rec} & \multicolumn{2}{c|}{FDSA} & \multicolumn{2}{c|}{CL4SRec} & \multicolumn{2}{c|}{DuoRec} & \multicolumn{2}{c|}{ICLRec} & \multicolumn{2}{c|}{MAERec} & \multicolumn{2}{c}{\textbf{\model}} \\
         \cmidrule{3-22}
         & & @3 & @5 & @3 & @5 & @3 & @5 & @3 & @5 & @3 & @5 & @3 & @5 & @3 & @5 & @3 & @5 & @3 & @5 & @3 & @5\\
         \midrule
         \multicolumn{22}{c}{Cross-Domain in Amazon Dataset} \\
         \midrule
         \multirow{3}{*}{Baby} & Hit@1 & \multicolumn{2}{c|}{.0019} & \multicolumn{2}{c|}{.0019} & \multicolumn{2}{c|}{.0025} & \multicolumn{2}{c|}{.0024} & \multicolumn{2}{c|}{.0025} & \multicolumn{2}{c|}{.0024} & \multicolumn{2}{c|}{\underline{.0025}} & \multicolumn{2}{c|}{.0010} & \multicolumn{2}{c|}{.0016} & \multicolumn{2}{c}{$\textbf{.0273}^{*}$} \\
         & Hit & .0051 &.0090 & .0060 & .0091 & .0062 & .0099 & .0061 & .0096 & \underline{.0064} & .0099 & .0060 & .0093 & .0061 & \underline{.0103} & .0046 & .0082 & .0055 & .0090 & $\textbf{.0281}^{*}$& $\textbf{.0283}^{*}$ \\
         & NDCG & .0036 &.0052 & .0043 & .0056 & .0046 & .0061 & .0045 & .0060 & \underline{.0047} & .0061 & .0045 & .0059 & .0046 & \underline{.0063} & .0031 & .0045 & .0038 & .0052 & $\textbf{.0277}^{*}$& $\textbf{.0279}^{*}$ \\
         \midrule
         \multirow{3}{*}{Games} & Hit@1 & \multicolumn{2}{c|}{.0031} & \multicolumn{2}{c|}{.0032} & \multicolumn{2}{c|}{.0032} & \multicolumn{2}{c|}{.0032} & \multicolumn{2}{c|}{.0026} & \multicolumn{2}{c|}{\underline{.0045}} & \multicolumn{2}{c|}{.0031} & \multicolumn{2}{c|}{.0026} & \multicolumn{2}{c|}{.0018} & \multicolumn{2}{c}{$\textbf{.0364}^{*}$} \\
         & Hit & .0079 &.0133 & .0093 & .0133 & .0093 & .0139 & .0082 & .0135 & .0084 & .0135 & \underline{.0112} & \underline{.0163} & .0078 & .0117 & .0060 & .0107 & .0078 & .0119 & $\textbf{.0374}^{*}$& $\textbf{.0376}^{*}$ \\
         & NDCG & .0058 &.0080 & .0066 & .0083 & .0066 & .0084 & .0061 & .0083 & .0059 & .0080 & \underline{.0082} & \underline{.0103} & .0058 & .0074 & .0046 & .0065 & .0052 & .0069 & $\textbf{.0370}^{*}$& $\textbf{.0371}^{*}$ \\
         \midrule
         \multirow{3}{*}{Office} & Hit@1 & \multicolumn{2}{c|}{.0021} & \multicolumn{2}{c|}{.0019} & \multicolumn{2}{c|}{.0021} & \multicolumn{2}{c|}{.0021} & \multicolumn{2}{c|}{.0021} & \multicolumn{2}{c|}{.0021} & \multicolumn{2}{c|}{\underline{.0022}} & \multicolumn{2}{c|}{.0021} & \multicolumn{2}{c|}{.0020} & \multicolumn{2}{c}{$\textbf{.0280}^{*}$} \\
         & Hit & .0042 &.0059 & .0043 & .0053 & .0040 & .0059 & .0031 & .0049 & .0042 & \underline{.0061} & \underline{.0043} & .0058 & .0040 & .0056 & .0038 & .0049 & .0036 & .0047 & $\textbf{.0293}^{*}$& $\textbf{.0299}^{*}$ \\
         & NDCG & .0033 &.0040 & .0033 & .0037 & .0032 & .0040 & .0027 & .0034 & .0033 & \underline{.0041} & \underline{.0034} & .0040 & .0032 & .0039 & .0031 & .0035 & .0029 & .0034 & $\textbf{.0287}^{*}$& $\textbf{.0290}^{*}$ \\
         \midrule
         \multicolumn{22}{c}{Cross-Domain on Different Platforms} \\
         \midrule
         \multirow{3}{*}{Yelp} & Hit@1 & \multicolumn{2}{c|}{.0011} & \multicolumn{2}{c|}{.0008} & \multicolumn{2}{c|}{.0007} & \multicolumn{2}{c|}{.0013} & \multicolumn{2}{c|}{.0009} & \multicolumn{2}{c|}{.0021} & \multicolumn{2}{c|}{.0019} & \multicolumn{2}{c|}{.0003} & \multicolumn{2}{c|}{\underline{.0029}} & \multicolumn{2}{c}{$\textbf{.0161}^{*}$} \\
         & Hit & .0023 &.0036 & .0022 & .0039 & .0026 & .0040 & .0037 & .0054 & .0025 & .0038 & .0056 & .0083 & .0043 & .0065 & .0027 & .0045 & \underline{.0079} & \underline{.0119} & $\textbf{.0163}^{*}$& $\textbf{.0166}^{*}$ \\
         & NDCG & .0018 &.0023 & .0016 & .0023 & .0018 & .0024 & .0026 & .0033 & .0018 & .0023 & .0040 & .0051 & .0033 & .0042 & .0016 & .0024 & \underline{.0057} & \underline{.0074} & $\textbf{.0162}^{*}$& $\textbf{.0163}^{*}$ \\
         \midrule
         \multirow{3}{*}{Washington} & Hit@1 & \multicolumn{2}{c|}{.0012} & \multicolumn{2}{c|}{.0016} & \multicolumn{2}{c|}{.0010} & \multicolumn{2}{c|}{.0027} & \multicolumn{2}{c|}{.0016} & \multicolumn{2}{c|}{.0032} & \multicolumn{2}{c|}{.0024} & \multicolumn{2}{c|}{.0017} & \multicolumn{2}{c|}{\underline{.0041}} & \multicolumn{2}{c}{$\textbf{.0122}^{*}$} \\
         & Hit & .0030 &.0043 & .0037 & .0056 & .0033 & .0046 & .0073 & .0114 & .0038 & .0053 & .0085 & .0127 & .0067 & .0105 & .0043 & .0069 & \underline{.0103} & \textbf{.0154} & $\textbf{.0128}^{*}$& \underline{.0130} \\
         & NDCG & .0022 &.0028 & .0028 & .0036 & .0023 & .0028 & .0053 & .0070 & .0029 & .0035 & .0063 & .0079 & .0049 & .0064 & .0032 & .0042 & \underline{.0076} & \underline{.0097} & $\textbf{.0126}^{*}$& $\textbf{.0127}^{*}$ \\
         \midrule
         \multirow{3}{*}{Steam} & Hit@1 & \multicolumn{2}{c|}{.0396} & \multicolumn{2}{c|}{.0484} & \multicolumn{2}{c|}{.0121} & \multicolumn{2}{c|}{.0757} & \multicolumn{2}{c|}{\underline{.1022}} & \multicolumn{2}{c|}{.0681} & \multicolumn{2}{c|}{.1011} & \multicolumn{2}{c|}{.0014} & \multicolumn{2}{c|}{.0727} & \multicolumn{2}{c}{$\textbf{.1237}^{*}$} \\
         & Hit & .0675 &.0830 & .0737 & .0881 & .0350 & .0538 & .1002 & .1161 & \underline{.1231} & \textbf{.1366} & .0986 & .1177 & .1223 & \underline{.1351} & .0433 & .0678 & .1018 & .1172 & $\textbf{.1246}^{*}$& .1253 \\
         & NDCG & .0557 &.0621 & .0630 & .0689 & .0254 & .0331 & .0898 & .0964 & \underline{.1143} & \underline{.1232} & .0858 & .0936 & .1134 & .1187 & .0257 & .0357 & .0897 & .0960 & $\textbf{.1242}^{*}$& $\textbf{.1245}^{*}$ \\
         \bottomrule
    \end{tabular}
    }
    \label{tab:exp_zero_shot}
    \vspace{-0.05in}
\end{table*}
Zero-shot recommendation represents a critical capability for real-world deployment scenarios where new domains emerge without sufficient training data. To rigorously evaluate this challenging setting, we comprehensively assess \model's ability to make effective recommendations without any domain-specific fine-tuning across six diverse datasets spanning multiple recommendation contexts. For robust comparative analysis, we systematically benchmark against traditional sequence recommenders trained on 10\% of target domain data, as these methods fundamentally require ID-based representations and cannot operate in a true zero-shot paradigm. Table~\ref{tab:exp_zero_shot} presents these extensive comparative results, revealing key insights about cross-domain generalization capabilities:

\noindent $\bullet$ \textbf{i) Superior Zero-shot Performance}. \model\ demonstrates exceptional zero-shot prediction capabilities across evaluated domains, significantly outperforming traditional sequential recommenders even when they have access to 10\% of domain-specific data. This remarkable generalizability stems from our \textit{Unified Item Tokenization} mechanism (Section 2.1), which compresses semantic embeddings from disparate domains into cohesive token representations using our Finite Scalar Quantization (FSQ) scheme. By mapping diverse item descriptions into a standardized token space, our model addresses the semantic heterogeneity challenge that typically impedes cross-domain transfer.

\noindent $\bullet$ \textbf{ii) Enhanced Cross-domain Generalization}.
Our \textit{Universal Recommendation Modeling} architecture (Section 2.2) combines bidirectional attention for intra-item token processing with causal attention for sequence modeling, creating a hybrid mechanism that preserves both item coherence and sequential dependencies. This innovation allows \model\ to capture complex interaction patterns transcending domain boundaries, contributing significantly to cross-domain adaptability. Even advanced baselines incorporating auxiliary text features (\eg, GRU4RecF, FDSA) or employing sophisticated self-supervised learning techniques (\eg, CL4SRec, DuoRec, ICLRec, MAERec) fail to match \model's generalization capabilities without this critical attention design.

\noindent $\bullet$ \textbf{iii) Effective Knowledge Transfer}.
\model's exceptional cross-domain performance is enhanced by our auxiliary \textit{Semantic Feature Integration} (Section 2.2.2), which counteracts information loss during quantization by preserving critical semantic nuances that conventional approaches discard. This dual-stream embedding approach--combining discrete tokens with continuous semantic features--enables more effective knowledge transfer across domains. Existing sequence recommenders, which fundamentally depend on ID embeddings and substantial in-domain training data, struggle in cross-domain scenarios even with access to 10\% of domain-specific data, highlighting the effectiveness of our semantic-preserving architecture in addressing fundamental limitations of ID-based paradigms.

\subsection{Real-World Deployment and Industrial Validation}
\begin{figure}[h]
    \centering
    \begin{minipage}{0.48\textwidth}
        \centering
        \includegraphics[width=\textwidth]{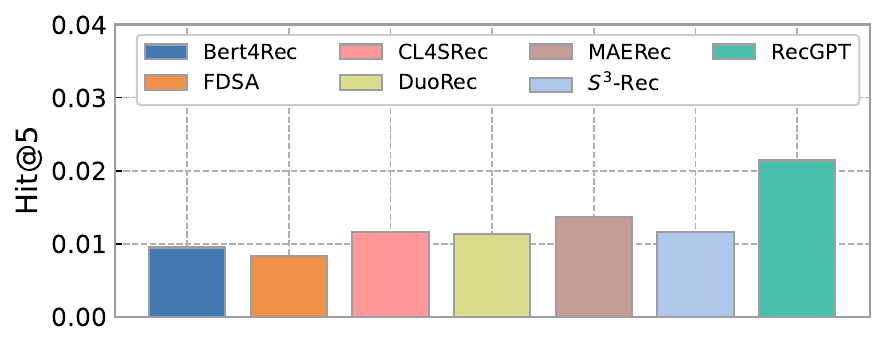}
        \caption*{(a) Performance w. Hit Rate}
    \end{minipage}
    \hfill
    \begin{minipage}{0.48\textwidth}
        \centering
        \includegraphics[width=\textwidth]{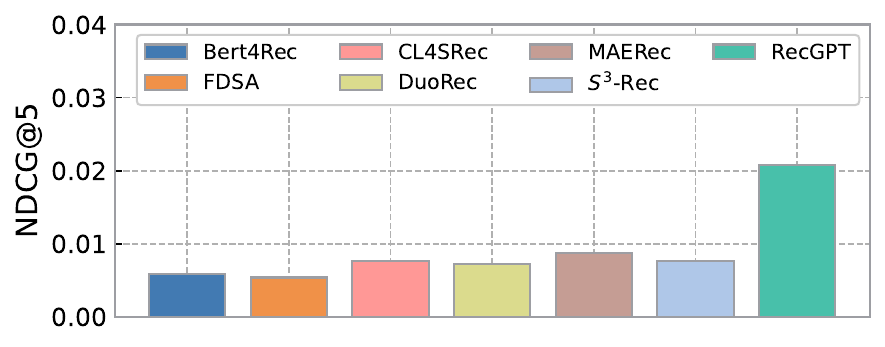}
        \caption*{(b) Performance w. NDCG}
    \end{minipage}
    \caption{Performance on Industrial Dataset. Zero-shot comparison of RecGPT against baselines on a production news platform. Results show RecGPT's superior (a) Hit Rate and (b) NDCG performance.}
    \label{fig:figure_exp_industrial}
\end{figure}
We have successfully integrated \model into a production-scale industrial recommendation system serving millions of users daily. Our offline feature generation pipeline—with scheduled embedding updates for active users and cold-start items—demonstrates both computational efficiency and practical viability in high-stakes commercial environments. This real-world deployment conclusively validates that our foundation model can consistently meet the stringent latency, throughput, and reliability requirements of modern recommendation platforms.

To rigorously assess \model's effectiveness in authentic commercial settings, we conducted extensive experiments on a large-scale proprietary dataset (labeled ``Industrial'') from a major online content platform with millions of active users. This dataset represents real-world news consumption patterns and contains 163,385 unique users, 455,372 distinct content items, and 999,140 interaction records—including user identities, consumed articles, content titles, and precise temporal information. All baseline configurations match those specified in Section~\ref{sec:exp_zero_shot} to ensure fair comparison.

As Figure~\ref{fig:figure_exp_industrial} shows, \model\ consistently outperforms all competitive baselines across key recommendation metrics on this industrial dataset. These results demonstrate that our unified tokenization approach effectively transfers knowledge across domains, even in complex environments with heterogeneous content types and diverse user behaviors. This superior performance stems from our model's ability to leverage extensive pre-training data to capture generalizable preference patterns that remain robust when applied to previously unseen commercial domains.

\subsection{Cold-Start Recommendation}

\begin{table*}[h]
    \centering
    \caption{Cold-start performance comparison across recommendation models. Results highlight \textbf{bold} (best) and \underline{underlined} (second-best) performance metrics. Statistical significance (p < 0.05) is denoted by * indicating meaningful improvements over all competing baseline approaches.}
    \scriptsize
    \vspace{-0.05in}
    \setlength{\tabcolsep}{0.25mm}
    {
    \begin{tabular}{c| l |c c|c c| c c| c c| c c| cc|cc|cc|cc|cc}
        \toprule
         \multirow{2}{*}{Data} & \multirow{2}{*}{Metric} & \multicolumn{2}{c|}{GRU4Rec} & \multicolumn{2}{c|}{GRU4RecF} & \multicolumn{2}{c|}{Caser} & \multicolumn{2}{c|}{BERT4Rec} & \multicolumn{2}{c|}{FDSA} & \multicolumn{2}{c|}{CL4SRec} & \multicolumn{2}{c|}{DuoRec} & \multicolumn{2}{c|}{ICLRec} & \multicolumn{2}{c|}{MAERec} & \multicolumn{2}{c}{\textbf{\model}} \\
         \cmidrule{3-22}
         & & @3 & @5 & @3 & @5 & @3 & @5 & @3 & @5 & @3 & @5 & @3 & @5 & @3 & @5 & @3 & @5 & @3 & @5 & @3 & @5\\
         \midrule
         \multirow{3}{*}{Baby} & Hit@1 & \multicolumn{2}{c|}{.0030} & \multicolumn{2}{c|}{.0030} & \multicolumn{2}{c|}{.0029} & \multicolumn{2}{c|}{.0030} & \multicolumn{2}{c|}{.0064} & \multicolumn{2}{c|}{.0092} & \multicolumn{2}{c|}{\underline{.0099}} & \multicolumn{2}{c|}{.0011} & \multicolumn{2}{c|}{.0062} & \multicolumn{2}{c}{$\textbf{.0165}^{*}$} \\
         & Hit & .0090 &.0123 & .0080 & .0122 & .0086  & .0124 & .0085 & .0120 & .0130 & .0168 & .0159 & .0193 & \underline{.0164} & \textbf{.0204} & .0057 & .0092 & .0124 & .0160 & $\textbf{.0171}^{*}$& \underline{.0172} \\
         & NDCG & .0063 &.0077 & .0058 & .0075 & .0061 & .0077 & .0061 & .0075 & .0101 & .0117 & .0130 & .0144 & \underline{.0136} & \underline{.0153} & .0038 & .0052 & .0098 & .0112 & $\textbf{.0168}^{*}$& $\textbf{.0169}^{*}$ \\
         \midrule
         \multirow{3}{*}{Office} & Hit@1 & \multicolumn{2}{c|}{.0020} & \multicolumn{2}{c|}{.0028} & \multicolumn{2}{c|}{.0024} & \multicolumn{2}{c|}{.0022} & \multicolumn{2}{c|}{.0088} & \multicolumn{2}{c|}{.0083} & \multicolumn{2}{c|}{\underline{.0096}} & \multicolumn{2}{c|}{.0049} & \multicolumn{2}{c|}{.0085} & \multicolumn{2}{c}{$\textbf{.0188}^{*}$} \\
         & Hit & .0053 &.0070 & .0060 & .0086 & .0045 & .0063 & .0041 & .0058 & .0149 & .0189 & .0150 & .0186 & \underline{.0172} & \textbf{.0207} & .0080 & .0096 & .0157 & .0189 & $\textbf{.0201}^{*}$& \underline{.0204} \\
         & NDCG & .0039 &.0046 & .0046 & .0057 & .0036 & .0043 & .0033 & .0040 & .0123 & .0140 & .0122 & .0137 & \underline{.0140} & \underline{.0154} & .0067 & .0074 & .0127 & .0140 & $\textbf{.0195}^{*}$& $\textbf{.0197}^{*}$ \\
         \midrule
         \multirow{3}{*}{Yelp} & Hit@1 & \multicolumn{2}{c|}{.0050} & \multicolumn{2}{c|}{.0049} & \multicolumn{2}{c|}{.0032} & \multicolumn{2}{c|}{.0023} & \multicolumn{2}{c|}{.0061} & \multicolumn{2}{c|}{.0057} & \multicolumn{2}{c|}{\underline{.0063}} & \multicolumn{2}{c|}{.0003} & \multicolumn{2}{c|}{.0057} & \multicolumn{2}{c}{$\textbf{.0126}^{*}$} \\
         & Hit & .0120 &.0164 & .0125 & .0180 & .0084 & .0129 & .0069 & .0107 & .0142 & .0219 & \textbf{.0149} & \textbf{.0221} & \underline{.0147} & \underline{.0220} & .0059 & .0102 & .0133 & .0196 & .0128& .0131 \\
         & NDCG & .0090 &.0108 & .0092 & .0115 & .0061 & .0080 & .0049 & .0065 & .0107 & \underline{.0139} & .0109 & .0138 & \underline{.0111} & \textbf{.0141} & .0035 & .0053 & .0101 & .0127 & $\textbf{.0127}^{*}$& .0128 \\
         \bottomrule
    \end{tabular}
    }
    \label{tab:exp_cold_start}
\end{table*}

Cold-start scenarios--where users have minimal historical interactions--represent one of the most persistent challenges in recommendation systems. To rigorously evaluate \model's effectiveness in addressing this limitation, we designed a controlled experiment that precisely simulates real-world cold-start conditions. For each user in our evaluation datasets, we randomly truncated their interaction sequences to contain only 1-3 items, using the subsequent interaction as the prediction target. This approach creates a realistic approximation of new users with limited platform engagement history.

As shown in Table~\ref{tab:exp_cold_start}, \model\ demonstrates remarkable performance advantages over specialized baselines across multiple metrics and domains, despite these challenging conditions. While competing methods—even when trained on the complete dataset with explicit cold-start optimization strategies—struggle to extract meaningful preference signals from such sparse interaction histories, \model\ excels through its foundation model architecture. This superior performance stems from \model's extensive pre-training across diverse recommendation contexts, which enables it to recognize generalizable preference patterns even from minimal user data.

The results highlight an advantage of our \model: by leveraging knowledge transferred from large-scale pre-training rather than relying solely on user-specific interactions, \model can effectively address the cold-start problem without requiring specialized adaptation techniques or extensive training. This capability has significant practical implications for recommendation systems, potentially eliminating the traditional trade-off between recommendation quality and new user experience.

\subsection{Ablation Study: Analyzing Component Contributions}

\begin{table}[h]
    \centering
    \footnotesize
    \caption{Ablation Study Results of \model.}
    \setlength{\tabcolsep}{0.5mm}
    {
    \begin{tabular}{l | c c c c c c}
        \toprule
        \multirow{2}{*}{\textbf{Variants}} & \multicolumn{2}{c}{\textbf{Baby}} & \multicolumn{2}{c}{\textbf{Office}} & \multicolumn{2}{c}{\textbf{Yelp}} \\
        \cmidrule(lr){2-3}
        \cmidrule(lr){4-5}
        \cmidrule(lr){6-7}
        & Hit@5 & NDCG@5 & Hit@5 & NDCG@5 & Hit@5 & NDCG@5 \\
        \midrule
        w/o FSQ & 0.0178 & 0.0177 & 0.0167 & 0.0166 & 0.0139 & 0.0138 \\
        w/o Bidir & 0.0279 & 0.0275 & 0.0288 & 0.0283 & 0.0162 & 0.0158 \\
        w/o Aux & 0.0191 & 0.0189 & 0.0205 & 0.0200 & 0.0075 & 0.0065 \\
        w/o Pref & 0.0282 & 0.0274 & 0.0281 & 0.0280 & 0.0162 & 0.0161 \\
        \midrule
        \textbf{\model} & \textbf{0.0283} & \textbf{0.0279} & \textbf{0.0299} & \textbf{0.0290} & \textbf{0.0166} & \textbf{0.0163} \\
        \bottomrule
    \end{tabular}
    }
    \label{tab:ablation}
    \vspace{-0.1in}
\end{table}

To systematically evaluate the architectural decisions underlying \model's exceptional cross-domain capabilities, we conducted comprehensive ablation experiments isolating the impact of each key component on zero-shot recommendation performance. Our analysis focuses on four critical variants:
\begin{itemize}[leftmargin=*]
\item \textbf{w/o FSQ}: replaces our FSQ-based unified item tokenizer with randomly assigned tokens, eliminating the semantic-preserving mapping that systematically transforms continuous embeddings into standardized discrete token sequences through the carefully optimized quantization process.

\textbf{w/o Bidir}: eliminates the bidirectional attention that enables comprehensive information sharing among tokens representing the same item. This variant relies on standard unidirectional causal attention for modeling both intra-item token relationships and inter-item sequential dependencies.

\item \textbf{w/o Aux}: removes the auxiliary semantic feature stream that complements discrete token embeddings with continuous semantic information. This ablation eliminates the dual embedding pathway that counteracts information loss from quantization through parallel representation learning.

\item \textbf{w/o Pre}: disables the Trie-based prefix enhancement technique during inference that constrains beam search to explore only token sequences leading to valid catalog items. This removes catalog-aware search optimization that improves computational efficiency and recommendation relevance.

\end{itemize}

\noindent Table~\ref{tab:ablation} presents our ablation results, revealing several critical insights about \model's architecture:

\noindent $\bullet$ \textbf{Unified Tokenization Impact.} The dramatic performance degradation observed in the \textit{w/o FSQ} variant demonstrates that our semantic-preserving tokenization is the cornerstone of RecGPT's cross-domain generalization capability. By transforming diverse textual item descriptions into a standardized token space, FSQ enables the model to identify universal preference patterns that transcend domain boundaries—a capability entirely absent when using random tokens.

\noindent $\bullet$ \textbf{Attention Mechanism Effectiveness.} The substantial performance drop in the \textit{w/o Bidir} variant confirms our hypothesis that standard unidirectional attention inadequately models intra-item token relationships. Our hybrid attention design, which combines bidirectional processing within items while maintaining causal attention between items, significantly enhances the model's ability to form coherent item representations while preserving sequential dependencies.

\noindent $\bullet$ \textbf{Dual-Stream Representation Benefits.} Results from the \textit{w/o Aux} ablation validate our approach of complementing discrete tokens with continuous semantic features. This dual-stream architecture effectively mitigates information loss during quantization, preserving critical semantic nuances that might otherwise be discarded. The auxiliary representations provide gradient pathways that enhance model training stability while enriching the semantic space available for preference modeling.

\subsection{Scaling Law Investigation of Recommendation Foundation Models}

\begin{figure}[h]
    \centering
    \captionsetup[subfigure]{labelformat=empty}
    \begin{subfigure}{0.32\columnwidth}
        \centering
        \includegraphics[width=\linewidth]{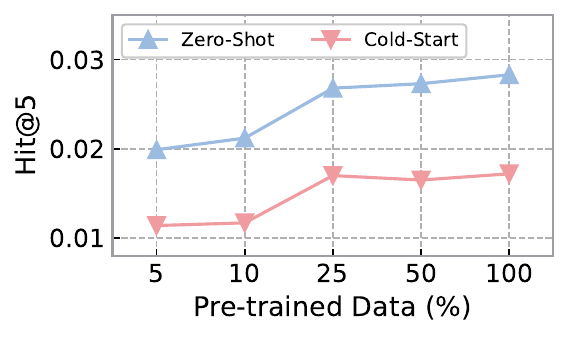}
    \end{subfigure}
    \hfill
    \begin{subfigure}{0.32\columnwidth}
        \centering
        \includegraphics[width=\linewidth]{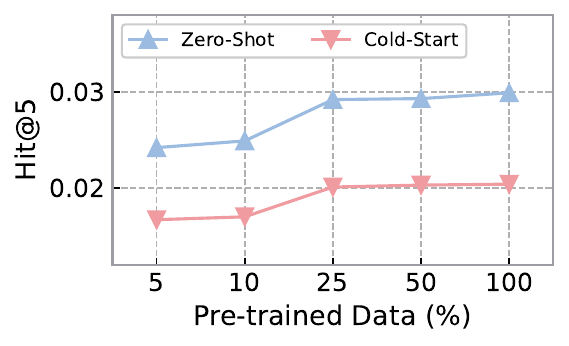}
    \end{subfigure}
    \hfill
    \begin{subfigure}{0.32\columnwidth}
        \centering
        \includegraphics[width=\linewidth]{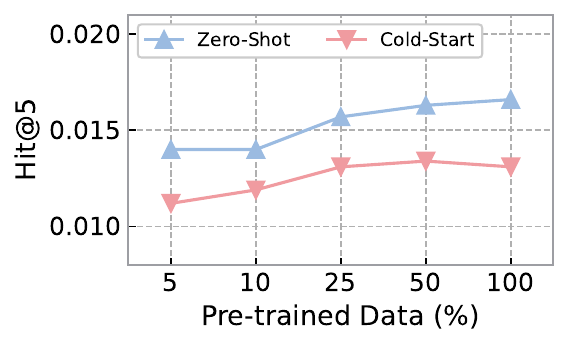}
    \end{subfigure}
    
    \begin{subfigure}{0.32\columnwidth}
        \centering
        \includegraphics[width=\linewidth]{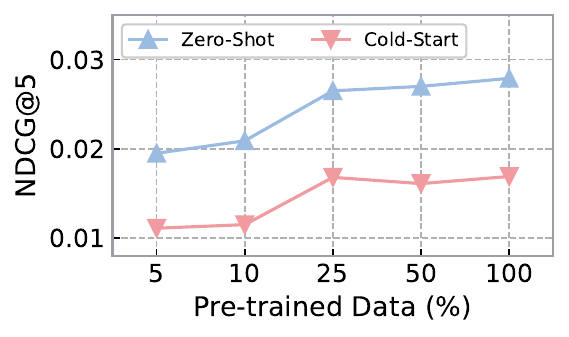}
        \subcaption{(a) Baby Dataset}
    \end{subfigure}
    \hfill
    \begin{subfigure}{0.32\columnwidth}
        \centering
        \includegraphics[width=\linewidth]{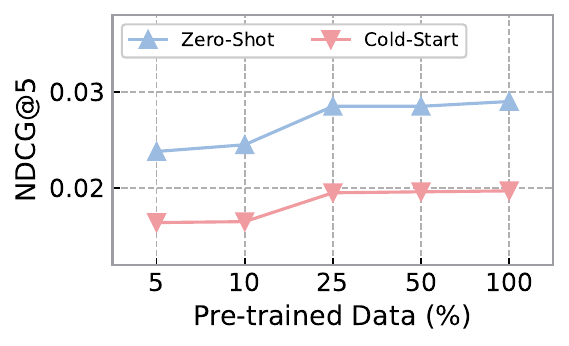}
        \subcaption{(b) Office Dataset}
    \end{subfigure}
    \hfill
    \begin{subfigure}{0.32\columnwidth}
        \centering
        \includegraphics[width=\linewidth]{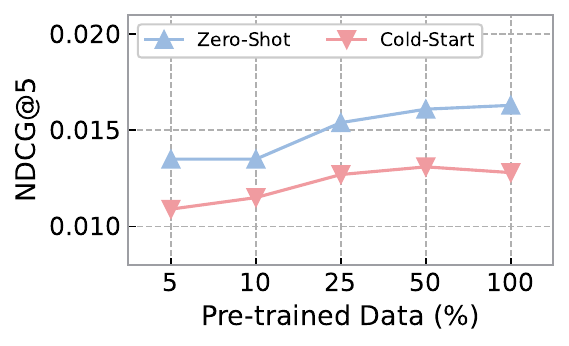}
        \subcaption{(c) Yelp Dataset}
    \end{subfigure}
    
    \caption{Performance \textit{w.r.t.} the volume of training data.}
    \label{fig:figure_exp_scale_pred}
\end{figure}

\begin{figure}[]
    \centering
    \begin{minipage}{0.45\columnwidth}
        \centering
        \includegraphics[width=\textwidth]{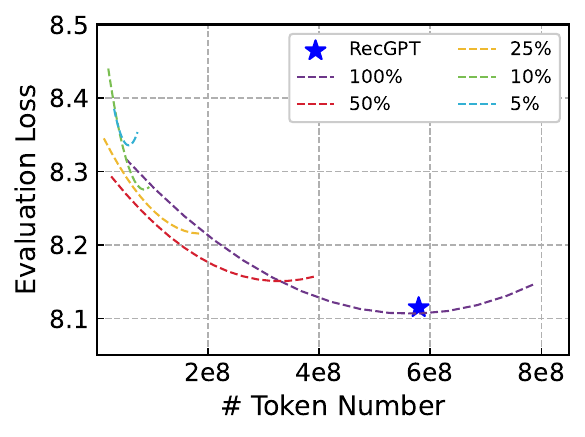}
    \end{minipage}\hfill
    \begin{minipage}{0.45\columnwidth}
        \centering
        \includegraphics[width=\textwidth]{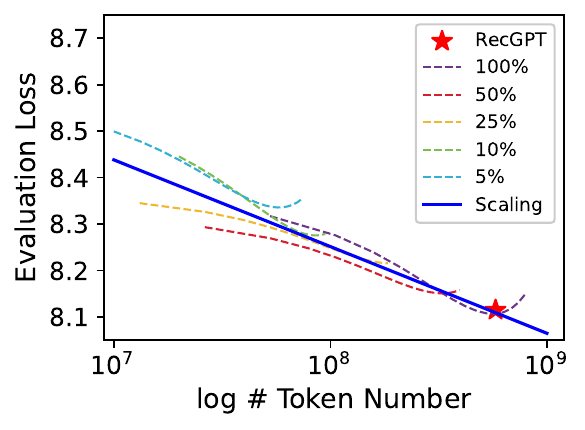}
    \end{minipage}
    \vspace{-0.1in}
    \caption{The scaling law of \model.}
    \label{fig:figure_exp_scale_law}
    \vspace{-0.1in}
\end{figure}

In this section, we investigate scaling laws for recommendation foundation models. Understanding how \model's performance scales with training data reveals opportunities for resource-efficient recommender systems while providing insights into cross-domain knowledge transfer. These scaling properties offer practitioners a framework for optimizing performance-computation trade-offs, potentially transforming how industry deploys recommendation systems at scale.

We trained five variants of \model\ using different proportions of the full training dataset (5\%, 10\%, 25\%, 50\%, and 100\%), while maintaining identical model architecture and hyperparameters. Each variant was trained until convergence, as indicated by stabilized evaluation loss. Figure~\ref{fig:figure_exp_scale_pred} presents the zero-shot recommendation performance across these variants, while Figure~\ref{fig:figure_exp_scale_law} illustrates the relationship between training tokens processed and evaluation loss.

Our investigation yields two key discoveries that are elaborated as follows.

\noindent \textbf{Emergent Generalization at Data Thresholds.} As shown in Figure~\ref{fig:figure_exp_scale_pred}, \model's zero-shot recommendation capabilities consistently strengthen with increased training data across all evaluated domains. Particularly noteworthy is the disproportionate performance jump between \model-10\% and \model-25\%, suggesting an emergent ability threshold~\cite{wei2022emergent} where the model develops substantially enhanced generalization capabilities once sufficient data diversity is available.

\noindent \textbf{LLM-Aligned Power-Law Scaling Properties.} \model\ demonstrates predictable scaling characteristics similar to those observed in LLMs~\cite{touvron2023llama}. By fitting a power-law curve to the evaluation losses of our model variants (Figure~\ref{fig:figure_exp_scale_law}), we can accurately predict the performance of the full model (indicated by \textcolor{red}{$\star$}). This confirms that \model, despite its recommendation-specific architecture, follows the established scaling laws that govern the recommendation foundation models.

These findings highlight a practical advantage for recommendation systems: rather than increasing model size (which impacts inference costs), expanding training data offers a more efficient path to performance improvements. The consistent scaling behavior suggests that \model's generalization capabilities could be further enhanced with additional training data without architectural changes—a valuable insight for production deployment scenarios where inference efficiency is paramount.

\subsection{Comparison with State-of-the-Art Pre-trained Recommenders}
This section benchmarks \model\ against leading pre-trained sequential recommenders to demonstrate its comparative advantages. We evaluate against diverse pre-training paradigms including self-supervised methods ($S^3$-Rec~\cite{zhou2020s3}), unified sequence models (UniSRec~\cite{hou2022towards}), vector-quantized approaches (VQ-Rec~\cite{hou2023learning}), transformer-based architectures (TIGER~\cite{rajput2023recommender}, RecFormer~\cite{li2023text}), and generative ID recommenders (IDGenRec~\cite{tan2024idgenrec}). For the ID-based $S^3$-Rec, we employ the same few-shot approach detailed in Section~\ref{sec:exp_zero_shot}, using 10\% of training data.

\begin{figure}[t]
    \centering
    \begin{minipage}{0.31\textwidth}
        \centering
        \includegraphics[width=\textwidth]{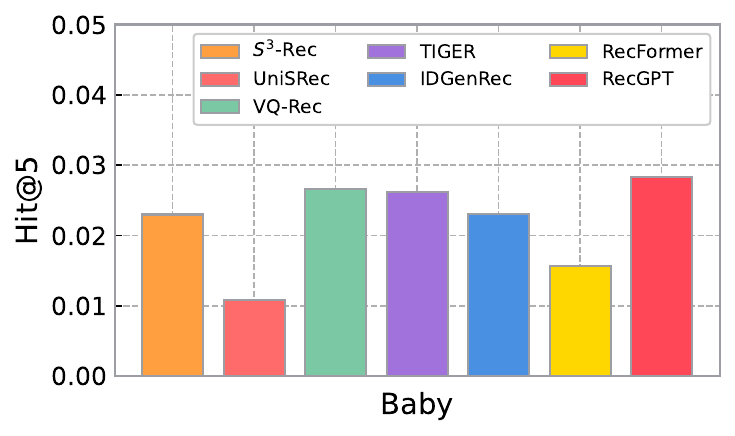}
    \end{minipage}\hfill
    \begin{minipage}{0.31\textwidth}
        \centering
        \includegraphics[width=\textwidth]{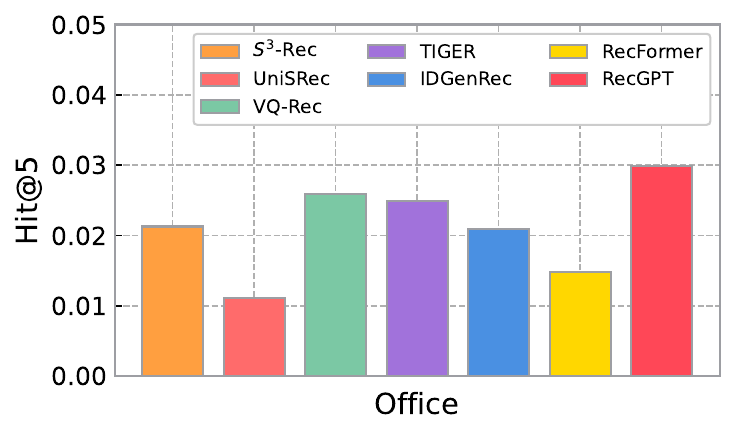}
    \end{minipage}\hfill
    \begin{minipage}{0.31\textwidth}
        \centering
        \includegraphics[width=\textwidth]{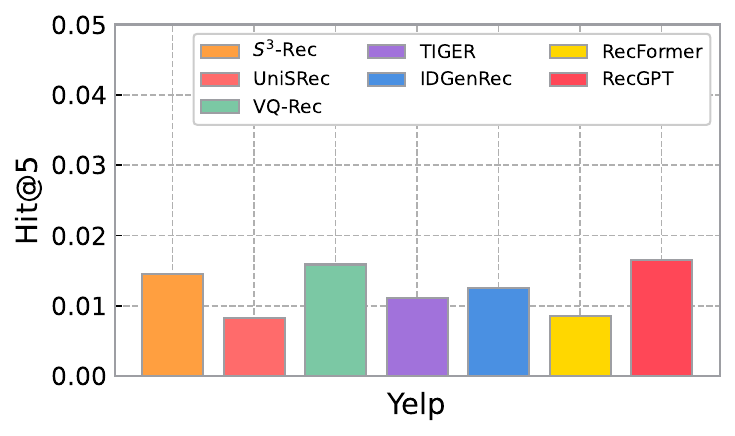}
    \end{minipage}
    
    \begin{minipage}{0.31\textwidth}
        \centering
        \includegraphics[width=\textwidth]{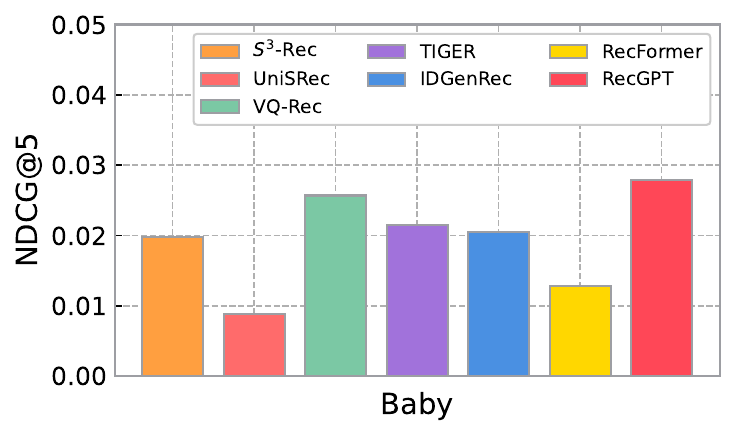}
    \end{minipage}\hfill
    \begin{minipage}{0.31\textwidth}
        \centering
        \includegraphics[width=\textwidth]{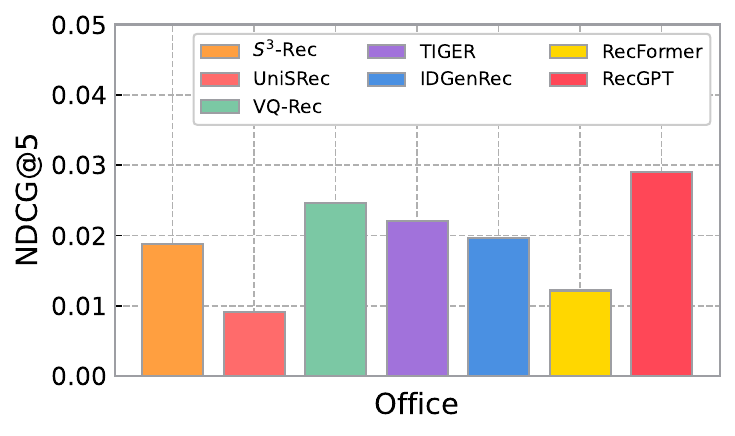}
    \end{minipage}\hfill
    \begin{minipage}{0.31\textwidth}
        \centering
        \includegraphics[width=\textwidth]{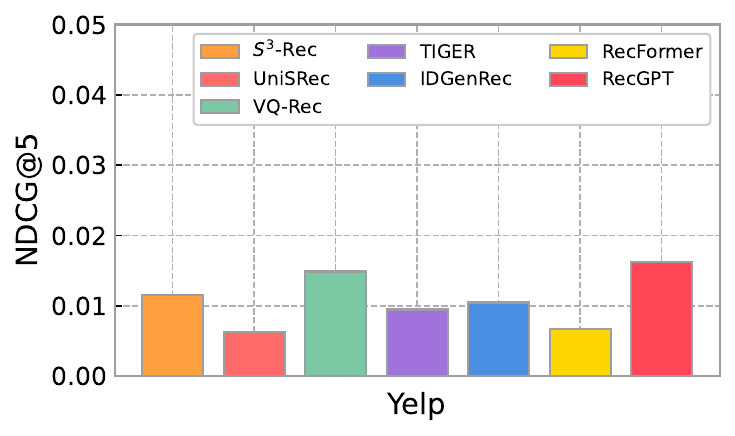}
    \end{minipage}
    \vspace{-0.05in}
    \caption{Performance comparison of pre-trained sequential recommenders across different datasets, measuring both ranking accuracy (HR@10, top) and quality (NDCG@10, bottom).}
    \label{fig:figure_exp_pre_train}
    \vspace{-0.1in}
\end{figure}

\noindent \textbf{Superior Performance Across Pre-training Paradigms.}
Figure~\ref{fig:figure_exp_pre_train} reveals RecGPT's consistent performance advantage over all other pre-trained sequential recommenders. While $S^3$-Rec significantly outperforms traditional methods from Table~\ref{tab:exp_zero_shot}—confirming the effectiveness of its pre-training tasks in mitigating data sparsity—it still falls short of RecGPT's capabilities. VQ-Rec emerges as the second-best performer, leveraging quantization techniques to enhance sequence generation and generalization. RecGPT's superior performance stems from three key advantages: (1) its decoder-only architecture that aligns with established scaling laws, (2) significantly larger pre-training data volume (ten million sequences versus one million for UniSRec and VQ-Rec, and three million for RecFormer), and (3) an autoregressive objective that better captures sequential dependencies.

\begin{figure}[t]
    \centering
    \begin{minipage}{0.16\columnwidth}
        \centering
        \includegraphics[width=\textwidth]{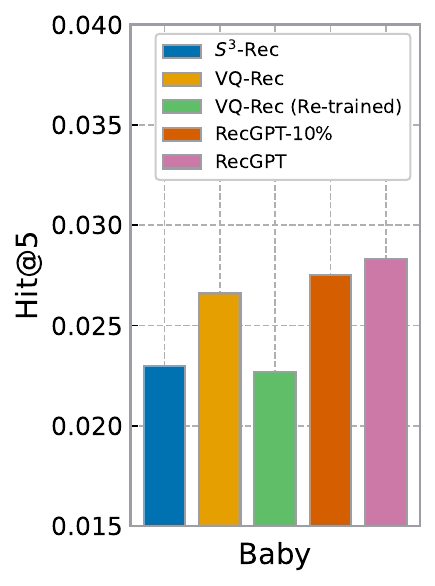}
    \end{minipage}\hfill
    \begin{minipage}{0.16\columnwidth}
        \centering
        \includegraphics[width=\textwidth]{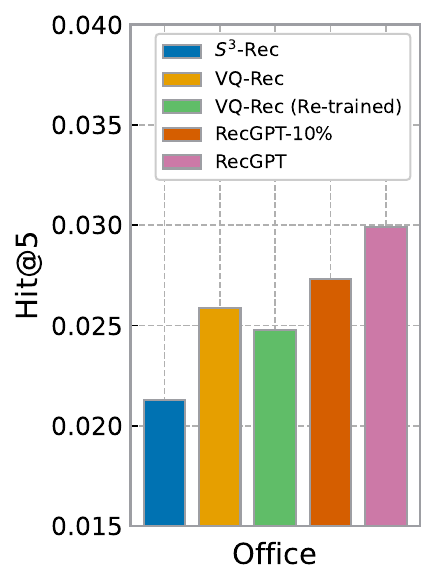}
    \end{minipage}\hfill
    \begin{minipage}{0.16\columnwidth}
        \centering
        \includegraphics[width=\textwidth]{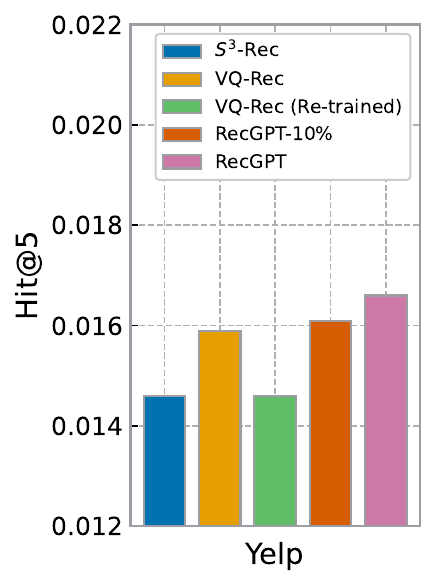}
    \end{minipage}
    \begin{minipage}{0.16\columnwidth}
        \centering
        \includegraphics[width=\textwidth]{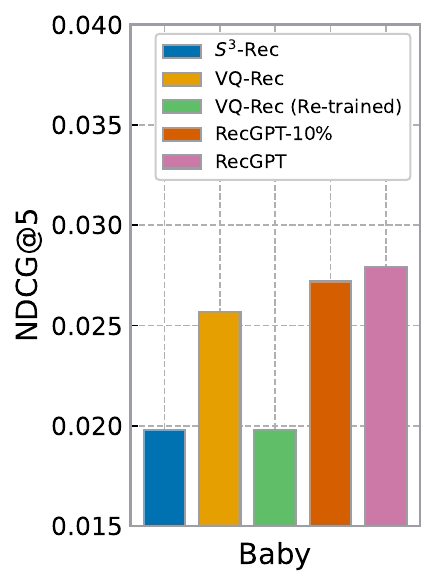}
    \end{minipage}\hfill
    \begin{minipage}{0.16\columnwidth}
        \centering
        \includegraphics[width=\textwidth]{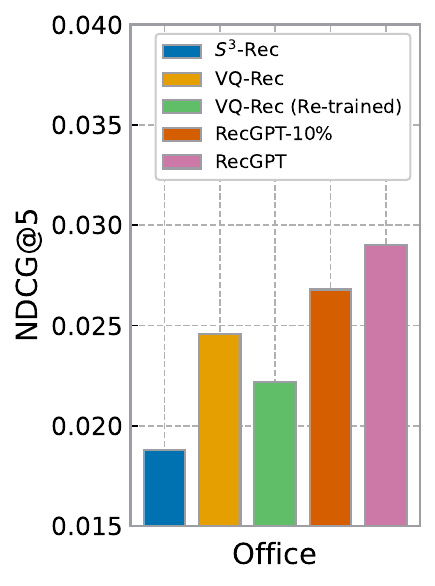}
    \end{minipage}\hfill
    \begin{minipage}{0.16\columnwidth}
        \centering
        \includegraphics[width=\textwidth]{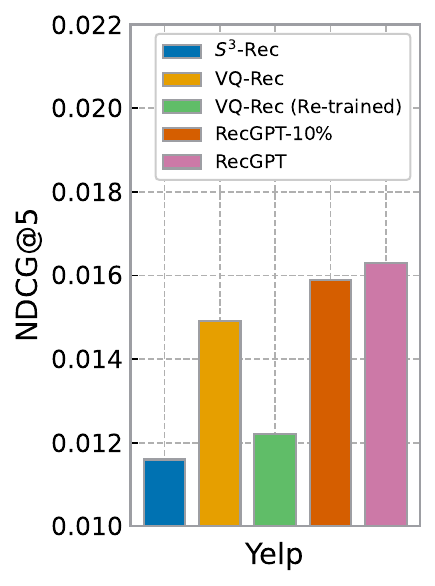}
    \end{minipage}
    \caption{Comparison isolating architectural benefits from data scale. \model-10\% outperforms VQ-Rec with equivalent training data, while VQ-Rec (Re-trained) shows performance degradation despite more data, confirming our model's architectural superiority.}
    \label{fig:figure_exp_vq-rec}
    \vspace{-0.1in}
\end{figure}

\noindent \textbf{Comparison with Vector-Quantized Approaches.}
To isolate the architectural benefits from data advantages, we conducted a controlled study with VQ-Rec—the strongest competitor from our initial comparison. We created two variants: \model-10\% (with reduced parameters and trained on 10\% of our dataset, matching VQ-Rec's training scale) and VQ-Rec (Re-trained) (using VQ-Rec's architecture but trained on our complete dataset). Figure~\ref{fig:figure_exp_vq-rec} demonstrates that even with equivalent training data, \model-10\% outperforms the original VQ-Rec. Surprisingly, VQ-Rec (Re-trained) shows performance degradation despite more training data, highlighting fundamental architectural limitations in its approach. This controlled experiment confirms that \model's performance advantage stems not merely from data scale but from its decoder-only architecture and autoregressive learning objective that better captures the nuances of sequential recommendation.

\section{Related Work}

\subsection{Sequential Recommendation}
Sequential recommendation has evolved significantly with the introduction of advanced neural architectures. Transformer-based models like SASRec~\cite{kang2018self} and BERT4Rec~\cite{sun2019bert4rec} revolutionized the field by leveraging self-attention mechanisms to capture complex sequential dependencies in user behavior. These architectures offered substantial improvements in both recommendation quality and computational scalability compared to earlier RNN-based approaches.

The limitations of existing ID-based embeddings~\cite{yuan2023go}—particularly their inability to generalize across domains—have inspired research into more transferable representations. Models such as UniSRec~\cite{hou2022towards} and VQ-Rec~\cite{hou2023learning} incorporate multi-modal item features to create more generalizable representations that can transfer knowledge across different recommendation scenarios. Concurrently, researchers have explored integrating large language models into recommendation systems~\cite{geng2022recommendation, zhang2023recommendation}, leveraging their semantic understanding capabilities to create text-based unified recommenders. Despite these advances, existing approaches still struggle with zero-shot generalization and typically require substantial domain-specific training data to perform effectively. Our work directly addresses these fundamental limitations by developing a foundation model approach that enables robust cross-domain generalization without domain-specific fine-tuning.

\subsection{Generative Recommender Systems}
Generative recommendation represents a paradigm shift where items are encoded as sequences of discrete tokens--often called semantic IDs~\cite{rajput2023recommender}--rather than continuous embeddings. This tokenization creates a more expressive and memory-efficient representation space by combining various discrete tokens to represent complex item characteristics~\cite{van2017neural}. Following principles established in language modeling~\cite{radford2019language, brown2020language}, generative recommenders process sequences of item tokens and autoregressively predict subsequent tokens, which are decoded into concrete item recommendations. Early implementations like TIGER~\cite{rajput2023recommender} employed residual quantization techniques to derive semantic codes, while subsequent research has focused on enhancing tokenization quality through improved quantization mechanisms~\cite{petrov2023generative, wang2024enhanced}. Notable advances include contrastive quantization~\cite{zhu2024cost} and learnable tokenizers~\cite{wang2024learnable, liu2024end} that adaptively optimize token representations for recommendation tasks.


\section{Conclusion, Limitation, and Future Work}
\model\ proposed in this work explores foundation models for recommender systems, built and trained from the ground up to address fundamental limitations in traditional approaches. Through our developed unified item tokenization and hybrid attention mechanisms, the model achieves exceptional zero-shot generalization capabilities across diverse recommendation domains without requiring fine-tuning. Comprehensive evaluations across public benchmarks and industrial deployment demonstrate \model's performance advantages over state-of-the-art methods in challenging scenarios including cross-domain recommendation, cold-start conditions, and production-scale environments.

Based on our industrial deployment experience with \model, we can discuss the limitations and future directions of recommendation foundation models from three key dimensions:

\noindent \textbf{Recommendation-Specific Challenges}: Unlike pure language understanding, the domain of recommender systems faces unique complexities. The intricate nature of human preferences, diversity of recommendation contexts, and heterogeneous input features across numerical, textual, visual, and audio modalities form bottlenecks for recommendation foundation models. While NLP benefits from text as a unified representational foundation, integrating these diverse modalities remains an ongoing research frontier. Future work should focus on developing unified architectures capable of seamlessly processing multimodal information to capture the multidimensional nature of user preferences.

\noindent \textbf{LLM Knowledge Integration with Efficiency Constraints}: Current recommendation foundation models have not fully leveraged rich world knowledge embedded in LLMs. Future models can combine LLMs' extensive knowledge with recommendation foundation models' specialized interaction patterns to better serve users. However, the challenge lies in addressing efficiency limitations when deploying LLM-scale models in industrial recommendation settings, where latency and computational constraints are particularly stringent. Developing lightweight knowledge distillation techniques or efficient cross-model integration approaches will be crucial for practical applications.

\noindent \textbf{Core Signal Integration and Adaptation}: While recommendation foundation models provide valuable complementary signals, particularly in cold-start scenarios, they cannot yet fully replace certain fundamental signals like core user profile information. Future research should explore how foundation models can better incorporate these essential signals while maintaining their generalization capabilities. Developing more sophisticated adaptation mechanisms that preserve the transfer learning benefits of foundation models while effectively leveraging domain-specific information represents a promising direction for creating truly comprehensive recommendation solutions.



\clearpage

\bibliographystyle{unsrtnat}
\bibliography{main}

\clearpage
\section{Appendix}

\subsection{Datasets}
\label{app:data}
The comprehensive evaluation framework encompasses diverse datasets spanning multiple domains and recommendation scenarios, enabling rigorous assessment of \model's cross-domain generalization capabilities. Dataset statistics are presented in Table~\ref{tab:dataset}, with detailed descriptions below.

\noindent $\bullet$ \textbf{Pre-training Data.} To establish robust cross-domain knowledge, we curate eleven diverse categories from Amazon review datasets~\cite{hou2024bridging}: \textit{All Beauty}, \textit{Books}, \textit{Clothing Shoes and Jewelry}, \textit{Electronics}, \textit{Health and Household}, \textit{Kindle Store}, \textit{Home and Kitchen}, \textit{Magazine Subscriptions}, \textit{Movies and TV}, \textit{Cell Phones and Accessories}, and \textit{Sports and Outdoors}. This heterogeneous collection provides comprehensive coverage of e-commerce recommendation patterns across distinct product domains.


\noindent $\bullet$ \textbf{Validation During Pre-training.} Three additional Amazon categories serve as held-out evaluation data during model development: \textit{Amazon Fashion}, \textit{Musical Instruments}, and \textit{Industrial and Scientific}. These datasets enable monitoring of generalization performance throughout the pre-training process without contaminating the final test evaluation.

\noindent $\bullet$ \textbf{Zero-Shot Evaluation Benchmark.} Our test suite comprises two complementary evaluation tracks. First, we assess same-platform generalization using three unseen Amazon categories (\textit{Baby Products}, \textit{Video Games}, and \textit{Office Products}) that share e-commerce characteristics with pre-training data but represent entirely new product domains. Second, we evaluate cross-platform generalization using three datasets from different recommendation contexts: \textbf{Yelp}\footnote{https://www.yelp.com/dataset}, containing user ratings on local business venues; \textbf{Washington}\cite{li2022uctopic, yan2023personalized}, featuring review information from Google Maps with rich metadata from Washington state; and \textbf{Steam}~\cite{kang2018self}, representing gaming preferences from Steam video game platform. This dual-track evaluation strategy comprehensively tests RecGPT's ability to generalize within e-commerce domains and across different recommendation paradigms.

\begin{table}[!htbp]
    \centering
    \small
    \caption{Statistics of experimental datasets used to evaluate \texttt{\model}'s performance.}
    \label{tab:dataset}
    \resizebox{0.5\textwidth}{!}{ 
    \setlength{\tabcolsep}{0.5mm}
    \begin{tabular}{l r r r r}
        \toprule
        \textbf{Datasets} & \textbf{\#Users} & \textbf{\#Items} & \textbf{\#Interactions} & \textbf{\#Avg.Len.} \\
        \midrule
        \textbf{Pre-trained} & 12,472,073 & 15,491,643 & 131,657,450 & 10.56 \\
        - Beauty & 1,464 & 6,570 & 13,679 & 9.34 \\
        - Books & 1,091,587 & 2,978,216 & 13,859,969 & 12.69 \\
        - Clothing & 3,088,673 & 4,875,707 & 30,245,204 & 9.79 \\
        - Electronics & 1,692,840 & 1,128,480 & 16,248,100 & 9.59 \\
        - Health & 828,935 & 518,044 & 7,598,392 & 9.16 \\
        - Kindle & 902,107 & 1,221,419 & 16,242,839 & 18.01 \\
        - Kitchen & 3,096,330 & 2,742,128 & 30,758,013 & 9.93 \\
        - Magazine & 380 & 865 & 2,679 & 7.05 \\
        - Movies & 653,846 & 569,357 & 7,622,246 & 11.65 \\
        - Phones & 547,327 & 628,147 & 4,103,048 & 7.49 \\
        - Sports & 568,584 & 822,710 & 4,963,281 & 8.73 \\
        \midrule
        \textbf{Evaluation} & 184,674 & 377,186 & 1,615,405 & 8.75 \\
        - Fashion & 13,942 & 76,873 & 104,115 & 7.46 \\
        - Instruments & 76,012 & 116,447 & 711,607 & 9.36 \\
        - Scientific & 94,720 & 183,866 & 799,683 & 8.44 \\
        \midrule
        \textbf{Baby} & 184,851 & 123,537 & 1,551,060 & 8.39 \\
        \textbf{Games} & 117,742 & 83,137 & 1,030,529 & 8.75 \\
        \textbf{Office} & 333,744 & 363,786 & 2,735,472 & 8.19 \\
        \midrule
        \textbf{Yelp} & 287,116 & 148,523 & 4,392,168 & 15.29 \\
        \textbf{Washington} & 625,428 & 120,080 & 12,382,314 & 19.79 \\
        \textbf{Steam} & 334,594 & 15,066 & 4,214,640 & 12.59 \\
        \bottomrule
    \end{tabular}%
    }
\end{table}

\subsection{Compared Methods}
\label{app:baselines}
To rigorously evaluate \model's effectiveness, we establish a comprehensive baseline framework encompassing representative methods from four major research streams in sequential recommendation. This systematic comparison ensures thorough assessment across diverse modeling paradigms.

\noindent $\bullet$ \textbf{RNN-based Sequential Models}. GRU4Rec~\cite{tan2016improved} pioneered session-based recommendation using GRU networks with ranking-based loss functions. GRU4RecF~\cite{hidasi2016parallel} enhanced this approach by incorporating both user click patterns and item features through deep learning architectures.

\noindent $\bullet$ \textbf{CNN-based Pattern Recognition}. Caser~\cite{tang2018personalized} introduces a novel perspective by transforming user interaction sequences into 2D image-like representations across temporal and latent dimensions, applying convolutional filters to capture local sequential patterns effectively.

\noindent $\bullet$ \textbf{Transformer-based Attention Mechanisms}. BERT4Rec~\cite{sun2019bert4rec} adapts the bidirectional encoder from BERT to sequential recommendation through Cloze task formulation. FDSA~\cite{zhang2019feature} leverages self-attention networks to identify complex patterns in both item and feature transitions.

\noindent $\bullet$ \textbf{Contrastive Learning Frameworks}. CL4SRec~\cite{xie2022contrastive} employs sequence-level augmentation strategies including item cropping, masking, and reordering. DuoRec~\cite{qiu2022contrastive} addresses representation degeneration through contrastive learning. ICLRec~\cite{chen2022intent} enhances recommendations by clustering and contrasting user intentions. MAERec~\cite{ye2023graph} dynamically distills global item transitional information for self-supervised augmentation, addressing label scarcity and noise in sequential recommendation.

\subsection{Implementation Details}
\label{app:implementation}

We implement our approach using the Transformers library from Hugging Face\footnote{https://huggingface.co}, a state-of-the-art platform for machine learning model development and deployment. Our implementation leverages several key architectural components designed to optimize sequential recommendation performance.

\noindent $\bullet$ \textbf{Item Representation Strategy.} To capture rich semantic information, each item's embedding is partitioned into four distinct segments, representing each item through four specialized tokens. This multi-token approach enables nuanced characterization compared to single-token representations.

\noindent $\bullet$ \textbf{Model Architecture Configuration.} The model operates with a maximum input sequence length $T$ of 1,024 tokens and hidden dimension $d_{ar}$ of 768. Correspondingly, the position embedding matrix $E_{wpe}$ has dimensions $\mathbb{R}^{1025\times 768}$, accommodating the sequence length plus one additional position.

\noindent $\bullet$ \textbf{FSQ Quantization Setup.} Our Finite Scalar Quantization (FSQ) module employs $d_{fsq} = 5$ dimensions with corresponding quantization levels $L$ of [8, 8, 8, 6, 5]. This configuration discretizes each item embedding sub-vector into a comprehensive vocabulary space of 15,360 unique tokens ($8 \times 8 \times 8 \times 6 \times 5 = 15,360$), ultimately resulting in a token embedding table of shape $\mathbb{R}^{15360\times 768}$.

\noindent $\bullet$ \textbf{Decoder Implementation.} The multi-layer decoder adopts the well-established, proven GPT-2 transformer architecture with 3 carefully selected layers, strategically balancing robust model expressiveness with computational efficiency for practical real-time recommendation deployment.

\noindent $\bullet$ \textbf{Baseline Configuration}. For comprehensive baseline evaluation, we leverage two established open-source recommendation libraries: RecBole~\cite{zhao2021recbole} and SSLRec~\cite{ren2023sslrec}. To ensure fair comparison, we standardize optimization using Adam while conducting extensive hyperparameter tuning for each baseline. We implement early stopping with 10-epoch patience to prevent overfitting.


\noindent $\bullet$ \textbf{Computational Infrastructure}. Our primary experiments utilize four A100 40G GPUs for optimal efficiency. However, we demonstrate accessibility by verifying that all experiments remain feasible on a single RTX 3090 24G GPU, requiring only batch size adjustments and longer training times.


\subsection{Evaluation Settings}
\label{app:evaluation_setting}
We evaluate next item prediction performance using two well-established standard ranking metrics: Hit@\textit{N} and NDCG@\textit{N}, with \textit{N} values of 1, 3, and 5. These complementary metrics comprehensively measure recommendation accuracy and ranking quality at different cutoff positions. For all six datasets, we employ a standard 9:1 split where 90\% of user interaction sequences constitute the training set, while 10\% forms the test set. During evaluation, we rank each sequence's ground-truth next item against the complete item catalog, then report averaged scores across all test users.


\end{document}